# Are events absolute?

*Hervé* Zwirn[*]

Centre Borelli (ENS Paris Saclay, France) & IHPST (CNRS, Université Paris 1)
herve.zwirn@gmail.com

**Abstract.** The "Wigner's Friend" thought experiment stands as one of the most intellectually provocative and challenging conceptual puzzles in quantum mechanics. It compels us to confront profound questions concerning the fundamental nature of reality, the very act of observation, and the possible role that consciousness might play within the quantum measurement process. This article gives a general presentation, beginning with Eugene Wigner's seminal proposal of the original thought experiment. In this paper, we explore its initial implications, which shook the foundations of classical physics, and then progress to an examination of the recent theoretical advancements and the ingenious extended versions of the experiment. The recent versions seem to imply that it is no more possible to consider events as absolute.

## 1. Introduction

Quantum mechanics (QM), undoubtedly one of the most successful scientific theories ever devised, has fundamentally reshaped our comprehension of the universe at its most fundamental scales. From the behaviour of subatomic particles to the intricate workings of lasers and semiconductors, its predictive power is undeniable. Yet, despite its empirical triumphs, quantum mechanics remains shrouded in a veil of enigma, particularly when it comes to the perplexing role of observers in shaping what we perceive as reality. This enduring mystery lies at the heart of the "measurement problem" – a central unresolved issue in quantum theory.

Among the various thought experiments designed to illuminate these philosophical quandaries, the "Wigner's Friend" scenario, originally introduced Eugene Wigner in 1961, serves as a quintessential example [1]. This ingenious construct vividly illustrates how the act of observation, or perhaps even the mere potential for observation, profoundly influences the behaviour and very existence of quantum systems. By presenting a seemingly paradoxical situation involving multiple levels of observation, Wigner's Friend forces us to confront critical questions about the precise nature of measurement, the elusive mechanism behind the "collapse of the wave function," and the complex interplay between subjective experiences and objective realities. Is there a definitive moment when a quantum state ceases to be a probabilistic superposition and settles into a single, determinate outcome? And what, or who, triggers this transition?

The enduring relevance of Wigner's Friend extends far beyond its initial conceptualization. Recent advancements in quantum information theory, coupled with an increasing capacity for experimental manipulation of complex quantum states, have not only reignited interest in this classic paradox but have also enabled the formulation and, in some cases, partial realization of sophisticated "extended" versions. These modern extensions, often involving nested observers and intricate entanglement, promise to further probe the limits of quantum theory and challenge our most deeply held intuitions about the fabric of reality. This paper aims to provide a thorough exploration of the original Wigner's Friend thought experiment and delve into its contemporary extensions, offering crucial insights into their far-reaching implications for both the foundational principles of physics and the enduring questions of philosophy. One weird consequence of these new "extended Wigner's Friend experiments" seem to be that events should not be considered as absolute.

## 2. The Original Wigner's Friend Thought Experiment

### 2.1 Setup and Concept

To fully grasp the profound implications of Wigner's Friend, it is essential to first understand its ingenious yet deceptively simple setup. The scenario involves two distinct observers, conventionally referred to as Wigner and his friend. The friend is situated inside a perfectly isolated, sealed laboratory, which is itself considered to be part of the quantum system. Within this isolated environment, the friend performs a quantum measurement on a quantum system. A common illustrative example is determining the spin state of a single electron, which, prior to any measurement, exists in a superposition of states –

---

[*] Corresponding author: herve.zwirn@gmail.com

simultaneously "spin-up" and "spin-down" in say the z-direction according to quantum mechanics.

According to the standard interpretation of quantum mechanics (specifically, the Copenhagen interpretation), when the friend makes their observation, the superposition is "collapsed." The friend registers a definite outcome – let's say, they definitively measure the electron to be "spin-up." From the friend's internal perspective, the measurement is complete, and the electron's state is now unequivocally defined.

However, the paradox emerges when we consider Wigner's perspective from *outside* the sealed laboratory. Since the friend and the experimental apparatus inside the lab are themselves physical systems, Wigner, applying the same universal laws of quantum mechanics, must treat the *entire* lab – including the friend, the electron, and the measurement device – as a single, large quantum system. From Wigner's vantage point, before he himself makes an observation of the lab, this combined system of "friend + electron + apparatus" exists in a superposition of states. This means Wigner would describe the lab as being in a state where the friend has measured "spin-up" *and* the friend has measured "spin-down" *simultaneously*, each correlated with the corresponding state of the electron.

The crucial question then becomes: When does the wave function collapse? Does it collapse when the friend makes their internal measurement within the lab, or does it only collapse when Wigner, the external observer, finally opens the lab and observes the entire system (including the friend's recorded outcome)? This discrepancy in perspective highlights a fundamental ambiguity in the standard formulation of quantum mechanics.

**2.2 Key Implications**

This seemingly straightforward thought experiment quickly unravels into a profound philosophical challenge. It directly questions the universality of quantum mechanics: if the theory applies to *all* physical systems, including conscious observers, then Wigner's description of the lab in superposition must be valid. Yet, the friend inside the lab experiences a definite reality. This leads to a puzzling dichotomy:

- Does the act of observation by the friend constitute a "measurement" sufficient to collapse the wave function for *everyone*, including Wigner? If so, what special property does a "friend" (or any observer) possess that causes this collapse?

- Or, is the friend themselves, along with their knowledge and experience, simply part of the larger quantum system as seen by Wigner? If this is the case, then consciousness itself would appear to be subject to the laws of superposition until observed externally, which runs counter to our everyday experience of a definite reality.

These inquiries extend far beyond mere academic curiosity; they delve into the very core of what quantum mechanics tells us about reality. Does it imply that quantum mechanics is an incomplete theory, lacking a crucial piece of information about the measurement process? Furthermore, the paradox leads to deeper inquiries into whether consciousness holds a unique or special role in the measurement process – a notion that has been both vigorously debated and widely dismissed within the physics community, often drawing criticism for introducing non-physical elements into scientific discourse. The Wigner's Friend experiment forces us to confront these uncomfortable questions head-on.

## 3. Philosophical and Interpretative Challenges

The Wigner's Friend thought experiment, which is very simple in its setup, acts as a powerful magnifying glass for the most persistent issue at the heart of quantum mechanics: the measurement problem. In particular, the highly contentious debate surrounding the role of consciousness.

**3.1 The Measurement Problem**

At its core, the measurement problem is the fundamental conundrum that quantum mechanics, in its standard formulation, fails to adequately define what constitutes a "measurement" or how, precisely, the wave function collapses [2]. According to the Schrödinger equation, quantum systems evolve deterministically and linearly, maintaining superpositions indefinitely until an observation is made. However, once a measurement occurs, this deterministic evolution abruptly ceases, and the wave function "collapses" into a single, definite state. The problem lies in the absence of a clear boundary or mechanism for this transition. When does "measurement" happen? Is it when a photon hits a detector, when an instrument records a result, or when a conscious mind perceives that result?

Wigner's Friend dramatically amplifies this ambiguity by introducing not one, but two nested levels of observation. From the friend's perspective, their measurement inside the lab immediately collapses the electron's superposition. They see a definite spin-up or spin-down. Yet, from Wigner's external viewpoint, the entire system (friend + electron + apparatus) remains in a superposition until he makes his own observation. This creates a logical inconsistency: is the collapse an event that occurs locally and immediately for the first observer, or is it postponed until the entire system is observed by a 'higher-level' observer?

This thought experiment forces us to confront whether the wave function collapse is a physical process – an objective, universally occurring phenomenon akin to a particle interaction – or an epistemic one, intrinsically linked to the acquisition of knowledge by an observer. If it's a physical process, why isn't it described by the same deterministic equations that govern other quantum evolutions? If it's epistemic, does it imply that reality itself is contingent on knowledge, or that knowledge itself can alter physical states? The lack of a clear, objective criterion for when and how collapse occurs remains arguably the single greatest unresolved issue in the foundations of quantum mechanics, and Wigner's Friend rings this conflict to the forefront with stark clarity.

**3.2 The Role of Consciousness**

Perhaps the most controversial implication Wigner himself considered was the potential role of consciousness in collapsing the wave function. Wigner, drawing inspiration from early interpretations like that of John von Neumann [3] or London and Bauer [4], speculated that it might be the conscious act of perception by the friend that instigates the collapse, transforming a quantum superposition into a definite classical reality. This concept aligns with interpretations sometimes dubbed "mind-body dualism" in the context of quantum mechanics, suggesting that consciousness is not merely a product of physical processes but might exert a causal influence on them.

This idea, however, remains profoundly contentious within the scientific community. Critics argue that introducing consciousness as a fundamental component of the measurement process introduces metaphysical elements into physics, making the theory untestable and potentially unfalsifiable. They point out that defining "consciousness" rigorously for scientific purposes is already a formidable challenge, let alone assigning it a unique physical role. Furthermore, if consciousness is required for collapse, what happens before conscious life emerged in the universe? Does it imply that quantum mechanics was incomplete or fundamentally different in the early universe?

Conversely, proponents, often few in number within mainstream physics but significant in philosophical discussions, see Wigner's proposition as an attempt to address a perceived inadequacy of purely materialist interpretations. They argue that if quantum mechanics is truly universal, applying to everything from electrons to observers, then the unique experience of a definite reality (rather than a superposition of realities) must be accounted for. For them, consciousness might be the very boundary where the quantum world transitions to the classical one we experience. While this view has largely fallen out of favour in mainstream quantum theory, it underscores the deep philosophical discomfort Wigner's Friend elicits regarding the interface between the objective quantum world and our subjective experience.

**3.3 Interpretational Diversity**

The profound ambiguities and paradoxes exposed by Wigner's Friend have naturally led to a rich tapestry of interpretations of quantum mechanics, each offering a different perspective on how to resolve the measurement problem and the observer's role. No single interpretation is universally accepted, and the Wigner's Friend experiment serves as a crucial thought experiment for testing the internal consistency and implications of each.

- Copenhagen Interpretation (CI): This is often considered the "standard" interpretation, developed by Niels Bohr and Werner Heisenberg. In the context of Wigner's Friend, CI generally posits that the wave function collapses *when the friend observes the system through a macroscopic device*. The friend's laboratory is considered part of the "classical" world where measurement outcomes are definite. The quantum state exists as a superposition until observed by a classical apparatus or a conscious observer. The core challenge for CI in the Wigner's Friend scenario is defining the precise "cut" between the quantum system and the classical measuring device/observer. Where exactly does the "classical" world begin and the quantum world end? The friend's perspective and Wigner's perspective then create a dilemma about when and where this "cut" truly lies.
- Many-Worlds Interpretation (MWI): Proposed by Hugh Everett III [5, 6], MWI offers a radically different solution: it postulates that the wave function *never* truly collapses. Instead, every time a quantum measurement is made, the universe "splits" or "branches" into multiple parallel universes, one for each possible outcome. In the Wigner's Friend scenario, when the friend measures the electron, the universe branches. In one branch, the friend measures

"spin-up," and in another, "spin-down." From Wigner's perspective, he too would eventually branch when he observes the lab, finding himself in one of these "worlds." MWI avoids the measurement problem by eliminating the need for collapse altogether, but it does so at the cost of positing an unfathomable number of parallel realities, each equally real. The discomfort often associated with MWI is first its ontological extravagance. Besides that, the status of probabilities is problematic since as all possible results happen, it becomes impossible to understand what the concept of probability means.
- Relational Quantum Mechanics (RQM): Developed by Carlo Rovelli [7], RQM takes a more nuanced approach, asserting that quantum states are not absolute properties of a system but are always relative to a specific observer. There is no "objective" quantum state of a system existing independently of any observation. For Wigner's Friend, this means the electron's state (and the friend's state) is in superposition *relative to Wigner*, but it is definite *relative to the friend.* Different observers can legitimately disagree on the state of the system because "facts" are relational. This interpretation resolves the paradox by dissolving the idea of a single, universal reality; instead, it proposes a network of relative facts. The problem is that RQM makes a confusion between correlation and measurement assessing that when two systems interact each one makes a measurement on the other, which is wrong. Besides that, the challenge for RQM is to explain how these relative facts coalesce into the shared classical reality we experience.
- Objective Collapse Theories (OCTs): These theories, such as those proposed by Ghirardi-Rimini-Weber (GRW) [8, 9] introduce modifications to the Schrödinger equation itself. They propose that collapse is a real, physical process that occurs spontaneously, independent of any observer. This collapse is typically triggered by certain physical conditions, such as the system reaching a certain size or complexity, or interacting with the environment in a specific way. For Wigner's Friend, an OCT would suggest that the wave function of the friend + electron system would collapse automatically due to its inherent complexity or interaction with the lab environment, even before Wigner observes it. The "friend's" observation merely records this objective collapse. OCTs aim to provide a concrete physical mechanism for collapse, but they typically require introducing new parameters or laws into quantum mechanics that are currently not supported by other experimental evidence.
- QBism: Some interpretations assume that the wave function does not represent the actual state of the system, but only refers to our knowledge of it. QBism is an interpretation of this type which was originally developed by Fuchs, Schack and Mermin [10]. QBists believe that the primitive concept of experience is the central subject of science. This is partly an instrumentalist position, since, for QBists, quantum mechanics is merely a tool enabling any agent to calculate his probabilistic expectations for his future experience from knowledge of the results of his past experience. Quantum mechanics therefore says nothing directly about the "outside world". A measurement (in the usual sense) is just a special case of what QBism calls experience, namely any action performed by an agent on his/her external world. A measurement does not reveal a pre-existing state of affairs, but creates a result for the agent. So the solution to the measurement problem is very simple: the agent's perception is the result of the experience. There is therefore no longer any ambiguity about the use of the reduction postulate, which is nothing more than the updating of the agent's state assignment on the basis of his or her experience. In particular, there is no measurement when there is no agent. In this case, the Wigner's friend experiment is no more puzzling.
- Convivial Solipsism (ConSol): This interpretation has been developed by the author [2, 11, 12, 13]. As in the Everett interpretation, there is no collapse and the wave function stays superposed. A measurement is not a physical action but the fact for an observer's perception to select one component of the superposed state. No observer has access to the perception to another observer. As a consequence there is no need for two observers neither to have gotten the same result nor to attribute the same state to the system. The results that each observer gets constitute their own reality that is not shared with the other observers. Hence there is no common reality but it is nevertheless possible to show that no disagreement can happen between different observers. The Wigner's friend situation is no more a problem. This interpretation solves the measurement problem and also avoids the need of non-local action which is supposed to exist in many other interpretations because of the violation of Bell's inequalities.

Each of these interpretations attempts to grapple with the Wigner's Friend paradox in its own way, highlighting the deep conceptual schisms within quantum foundations and the ongoing search for a coherent and universally accepted understanding of reality at its most fundamental level. The thought experiment thus serves as a critical benchmark against which the strengths and weaknesses of each interpretive framework can be evaluated.

## 4. Extended Versions of Wigner's Friend

The original Wigner's Friend thought experiment, for all its profound implications, largely remained a theoretical construct for decades. However, physicists have been able to conceive extended versions of Wigner's Friend. These modern iterations are not merely intellectual exercises; they are designed to push the boundaries of quantum mechanics, test its universal applicability, and probe the very consistency of its principles when applied to observers themselves. They introduce additional layers of complexity, often involving multiple observers, to amplify the paradoxes and highlight the divergences between different interpretations of quantum theory.

### 4.1 The "Extended Wigner's Friend" Scenarios

The power of the extended Wigner's Friend scenarios lies in their ability to create nested or entangled observational frameworks. Imagine the original setup, but now the friend has a "Wigner's Friend's Friend" inside the lab with the first friend, and also another "outside Wigner". This creates a multi-layered quantum system where each observer is a "friend" to the one outside them and a "Wigner" to the one inside.

A particularly insightful extended scenario involves two friends (Charlie and Debbie) and two Wigners (Alice and Bob) in separate, isolated laboratories. Charlie and Debbie share each one a particle from a pair of entangled particles.

- Charlie performs a measurement on one of the particle, say, spin-up or spin-down in the z-direction.
- Debbie measures the other entangled particle.
- Alice observes Charlie's lab, treating it as a superposition.
- Bob observes Debbie's lab, treating it also as a superposition.
- Crucially, these two external Wigners (Alice and Bob) can either perform a measurement on their respective labs, testing for a superposition of the labs themselves or open the lab and ask their friend for the result they got.

In such a nested arrangement, the system includes multiple layers of observation, with each observer potentially experiencing a different "reality" or state of affairs. Charlie experiences a definite electron spin. Debbie experiences their own definite outcome. However, Alice, observing Charlie's lab, might describe Charlie and the electron in a superposition. Similarly for Bob. When Alice and Bob then interact, or compare notes, their observations can lead to contradictions if a single objective reality is assumed. These extensions are specifically designed to test whether quantum mechanics can consistently describe observers within its own framework, leading to a kind of quantum meta-theory where observers themselves are treated as quantum objects.

### 4.2 Brukner's No-Go Theorem

One of the first theoretical developments emerging from the extended Wigner's Friend setups is Časlav Brukner's no-go theorem [14, 15]. Published in 2017, this theorem aims at demonstrating the following conclusion: if quantum mechanics is universally valid – meaning it applies to all systems, including observers – then there cannot be a single objective reality shared by all observers.

Brukner's theorem, building on the framework of extended Wigner's Friend scenarios, highlights that certain assumptions that seem intuitively reasonable (e.g., that facts are universally agreed upon) are incompatible with the predictions of quantum mechanics when applied consistently to observers. Specifically, it shows that if three conditions hold:

1. Locality: Measurements in one lab do not instantaneously affect distant labs.
2. No-Superdeterminism: Future choices of measurements are not pre-determined by hidden variables.
3. Universality of Quantum Mechanics: Quantum mechanics correctly describes all physical systems, including macroscopic ones and observers.

Then, the existence of objective facts (meaning facts that are true for all observers, regardless of their perspective) is contradicted in certain extended Wigner's Friend setups. In these scenarios, one observer might irrevocably commit to a specific measurement outcome, while another observer, encompassing the first, still describes the entire system (including the first observer's choice and outcome) as being in a superposition. This means observers, even if they communicate their results, might fundamentally disagree on whether a particular event (like a wave function collapse) has definitively occurred or what its outcome was. This has profound implications for our understanding of reality, suggesting a more observer-dependent or perspectival view of physical facts and challenging the classical notion of absolute scientific objectivity.

## 4.3 Other no-go theorems

Brukner's theorem can be contested trough the fact that it relies on too strong assumptions (namely to compare results that cannot be obtained in a single experiment). But following it, many other theorems have been proposed by Frauchiger and Renner [16], Pusey-Masanes [17], Omrod and Barrett [18], Bong et al. [19], Gao [20] etc... All these theorems go in the same direction: they assume very similar assumptions and derive a contradiction leading to the necessity to abandon at least one of the assumptions. Usually, it is the absoluteness of observed events (AOE) that is abandoned.

There is a very simple way to show how these theorems work. I show below a derivation coming from the excellent review paper [21].

Let's assume the following reasonable hypotheses:

1. Locality for observed events: Measurements in one lab do not instantaneously affect distant labs.
2. No-Superdeterminism: Future choices of measurements are not pre-determined by hidden variables.
3. Universality of Quantum Mechanics: Quantum mechanics correctly describes all physical systems, including macroscopic ones and observers.
4. Absoluteness of Observed Events (events are not relative to a particular observer but are true for all the observers)
5. A super Observer can undo a Measurement

The first four assumptions are intuitively acceptable. The fifth one is more difficult. It comes from the fact that if Quantum Mechanics is universal and there is no collapse, everything (including measurements) is described by unitary interactions that are invertible. So even a measurement can be undone by applying the inverse unitary interaction. Of course, this is possible only in principle because doing that really would imply a technology far beyond what is possible today and even in the far future. But we are here discussing of questions of principle.

As explained above, Charlie and Debbie are each one inside their own laboratory and share a particle coming from an entangled pair in the Hardy state:

$$\psi = \frac{1}{\sqrt{3}}\{|00\rangle + |01\rangle + |10\rangle\}$$

Charlie and Debby each measure their particle in the base $\{|0\rangle, |1\rangle\}$. They get the results c and d.

There is another possible base for measurement:

$$|+\rangle = \frac{1}{\sqrt{2}}\{|0\rangle + |1\rangle\}$$

$$|-\rangle = \frac{1}{\sqrt{2}}\{|0\rangle - |1\rangle\}$$

Then Alice can choose between asking Charlie which result he got (choice x=0) or undoing Charlie's measurement and measuring the particle in the base $\{|+\rangle, |-\rangle\}$. (choice x=1). She gets the result a.

Bob can choose between asking Debbie which result she got (choice y=0) or undoing Debbie's measurement and measuring the particle in the base $\{|+\rangle, |-\rangle\}$. (choice y=1). He gets the result b.

Now it is easy to verify the following equalities:

1. p(c=1, d= 1|x=0, y=0) = 0
2. p(c=0, b= -|x=0, y=1) = 0
3. p(a=-, d= 0|x=1, y=0) = 0
4. p(a=-, b= -|x=1, y=1) = 1/12 ≠ 0

Now, using the Locality for observed events we can deduce:

5. p(c=1, d= 1|x=1, y=1) = 0
6. p(c=0, b= -|x=1, y=1) = 0
7. p(a=-, d= 0|x=1, y=1) = 0
8. p(a=-, b= -|x=1, y=1) = 1/12 ≠ 0

To understand the reason why, let's consider the equality 1. Under the locality of observed events, the results that Charlie and Debbie get cannot depend on the choice of measurement of Alice and Bob that happen after Charlie and Debbie's measurements. Hence from the equality 1 we can deduce equality 5. The same reasoning proves the other four equalities. Equalities 5 to 8 concern result that can be gotten in a same run, so under the absoluteness of observed events, a, b, c, and d exist simultaneously.

Now: (a = -) => (d = 1) => (c=0) => (b = +)  hence p(a = -, b = -) = 0 which is in contradiction with equality 8.

The five hypotheses we have assumed lead to a contradiction. So we have to abandon at least one of them.

### 4.3 Experimental Realizations

For a long time, Wigner's Friend was considered a pure thought experiment, impossible to realize due to the practical challenges of keeping macroscopic observers in quantum superposition. However, recent advancements in quantum technologies have allowed for attempts to realize "experimental approximations" or "analogue realizations" of the Wigner's Friend scenario. While these experiments don't involve actual conscious human friends in superposition (which is still far beyond current capabilities), they use sophisticated quantum systems, typically entangled photons, to simulate the roles of the friend and the external Wigner.

These experiments involves a "friend" who is an automated quantum measurement device (e.g., a quantum computer or a system designed to perform a measurement and then retain that information in a coherent, superposed state). This "friend" performs a measurement on a quantum system (e.g., the polarization of a photon). Instead of immediately causing a definitive collapse for an external observer, the "friend's" interaction is reversible, meaning the information it gains is stored in a way that allows the combined "friend + system" state to remain in a superposition relative to the external "Wigner." The external "Wigner" then performs a measurement on the *entire composite system* (friend + original system), often an interference experiment, to verify if the superposition was maintained.

That is the case in a 2019 experiment by Massimiliano Proietti et al. [22], where entangled photons were used to simulate the Friends and the Wigners. According to these authors the results supported the idea that, from Wigner's perspective, the friend's measurement did not cause an irreversible collapse until Wigner himself performed his own measurement. The main issue with that kind of experiment is that it is not obvious at all that it is legitimate to use photons to play the role of an observer. It is even highly contestable that an interaction with a photon can be similar to a measurement. Hence, these experiments, even if they are beautiful experimental devices, are not conclusive.

### 4.4 A few reservations

Measurement: In the universal version of quantum mechanics where everything is a unitary process it is unclear what a measurement is.

Undoing a friend's measurement: If Wigner undoes the measurement done by Charlie, it becomes difficult to consider that the result of Charlie's measurement is still a fact.

Real experiences are totally out of reach: Experiences involving a macroscopic system like an observer would need a measurement apparatus much larger than the whole universe.

## 5. Implications of Extended Interpretations

The journey from the original Wigner's Friend thought experiment to its modern, extended versions, alongside the insights from theorems like those mentioned above, has profound implications that ripple across the landscape of physics and philosophy. These scenarios don't just confirm the strangeness of quantum mechanics; they amplify it, forcing us to fundamentally reconsider deeply ingrained classical intuitions about reality, causality, and the very nature of events. The consequences touch upon foundational debates, the role of observers, and even the practical aspects of quantum information theory.

### 5.1 Challenges to Classical Intuition

One of the most immediate and striking consequences of the extended Wigner's Friend scenarios is their direct challenge to classical intuition. Our everyday experience, informed by classical physics, leads us to assume a world where objects have definite properties regardless of observation, where events unfold in a universally agreed-upon sequence, and where a single, objective reality exists independent of any observer. The extended Wigner's Friend scenarios shatter these assumptions.

If we accept the assumption of universality of quantum mechanics according to which an external Wigner can describe an entire laboratory (including an internal "friend" who has already made a measurement) as being in a superposition, then the very notion of a definite, pre-existing reality becomes problematic. It forces us to reconsider the traditional boundary between the classical and quantum realms. Where does the fuzziness of the quantum world end and the definiteness of our macroscopic experience begin? These scenarios suggest that this boundary is not fixed or objective, but rather highly dependent on the observer's perspective and the scope of their observation. This radical shift demands a fundamental re-evaluation of concepts like realism (the belief that reality exists independently of our minds). They demonstrate that if quantum mechanics is truly universal, then the world we experience might be far less "real" in the classical sense than we intuitively believe. This is the picture that was adopted in Convivial Solipsism well before these no-go theorems which go in the same direction even though they rely on stronger assumptions that those made in ConSol.

## 5.2 Convivial Solipsism and Observer Dependence

The extended Wigner's Friend scenarios provide some of the strongest conceptual support for all the relational or perspectival interpretations such as Convivial Solipsism and its core tenet of observer dependence. As previously discussed, ConSol posits that quantum states are not absolute properties of a system, but are always defined relative to a specific observer. In essence, there is no "view from nowhere" in quantum mechanics; every quantum fact is a relation between a system and an observer.

In the multi-layered settings of extended Wigner's Friend, this relational aspect becomes exquisitely clear. Charlie's measurement outcome is definite for Charlie. Alice's description of Charlie's lab, however, treats Charlie and their measurement as being in a superposition, which is definite for Alice only after his observation. The tension between the "facts" held by different observers is precisely what ConSol addresses. It doesn't view this as a paradox or a contradiction, but rather as an inherent feature of reality: different observers simply have different, yet equally valid, sets of relative facts.

This interpretation suggests that the wave function is not an objective physical entity that collapses, but rather represents the set of an observer's potentialities about a system. The "collapse" then becomes a selection by the observer's perception of one among the many possibilities, not a physical process. The extended Wigner's Friend scenarios provide compelling arguments for this view by demonstrating that insisting on a single, absolute reality leads to inconsistencies, whereas embracing relativity of quantum facts offers a consistent, albeit counter-intuitive, framework.

## 5.3 Consequences for the Relativity of Events

Perhaps one of the most striking and counter-intuitive consequences highlighted by the extended Wigner's Friend scenarios is that events can no longer be considered absolute. Instead, they appear to become relative to each observer. In classical physics, an event (like a collision or a measurement outcome) happens at a specific point in space-time, and all observers, regardless of their motion (Lorentz transformations notwithstanding), agree that the event occurred and what its outcome was. This forms the basis of a shared, objective timeline.

However, in extended Wigner's Friend scenarios, two observers might fundamentally disagree on whether a particular quantum measurement has occurred or what its outcome was. For instance, Charlie inside the lab knows the electron's spin is definitively "up." But Alice, observing the entire lab from outside, might describe the electron and Charlie as being in a superposition of "up" and "down." From Alice's perspective, the "collapse" event for the electron has not yet occurred for the entire system he is observing. This means that an event that is a "fact" for one observer (the definite outcome for the friend) is merely a component of a superposition for another (the Wigner).

In ConSol nevertheless, there can be no disagreement between two observers because if they communicate about the result of a measurement they will agree on this result. This is because any communication between two observers must be understood as a measurement of one observer by the other. If Alice asks Bob which result he got, this is a measurement of Bob by Alice. Alice will hear Bob giving an answer that is conform to the result she got. But that does not mean that Bob will have gotten the same result from his standpoint. Saying that they both got the same result is a claim that can be expressed only from God's point of view which is forbidden in ConSol. A statement has to be made from the point of view of one unique observer. So it is not allowed to mention the result gotten by two different observers in one sentence. This relativity of events profoundly undermines the classical view of a shared, objective reality and introduces a deeply subjective layer to the sequence and nature of quantum occurrences. It necessitates a radical rethinking of how we define the reality.

If an event is not absolute, can causality be absolute? If information is not universally agreed upon, what does that mean for how knowledge is constructed in the quantum realm? This consequence is perhaps the most difficult for our classical minds to grasp, as it suggests a breakdown of a fundamental element of our experienced reality.

## 5.4 Impacts on Quantum Information Theory

Beyond the profound philosophical implications, extended Wigner's Friend scenarios raise also questions on quantum information theory on the nature of quantum computation and the role of measurement in quantum algorithms. If measurement outcomes are relative, what does that mean for the final state of a quantum computation? Do all branches in a Many-Worlds Interpretation contribute to the "answer" of a quantum computer? How does the "readout" process, which is essentially a measurement, affect the integrity and interpretation of quantum information? These thought experiments inspire new research into how information is fundamentally defined and processed in a multi-observer quantum universe. They help refine our understanding of information transfer, the limits of quantum error correction, and the foundational principles underlying future quantum technologies.

## 6. Philosophical Reflections

The enduring and evolving Wigner's Friend thought experiment, particularly its extended manifestations, transcends the realm of theoretical physics to delve deeply into fundamental philosophical question**s** that have puzzled humanity for millennia. These scenarios force us to confront not only the limits of our scientific theories but also the very nature of existence, knowledge, and our place within the cosmos. They compel a rigorous re-examination of our most basic assumptions about how reality is structured and how we come to know it.

### 6.1 Ontological Questions

Perhaps the most profound philosophical questions amplified by extended Wigner's Friend experiments are ontological in nature – questions about what truly *exists* or what it means for something to be "real." If, as these scenarios suggest, different observers can legitimately hold conflicting "facts" about the same system, particularly whether a quantum state has collapsed or not, then what does it mean for something to be "real"? Does reality exist independently of any observation, or is it fundamentally intertwined with the act of observation itself?

Classical physics strongly supported a view of objective realism: there is a single, objective reality out there, independent of human minds, and science's role is to discover its pre-existing properties. Wigner's Friend challenges this bedrock assumption. If the friend experiences a definite outcome while Wigner still describes Charlie and the system in a superposition, or even weirder, if in ConSol two observers can see different results, then whose reality is the "true" one? This forces us to consider whether reality is inherently subjective (meaning dependent on the individual observer), or if there is an underlying objective reality that somehow transcends or is hidden beneath our observational capacities. Some interpretations lean towards a multiplicity of realities (Many-Worlds), others towards relative realities (QBism, ConSol), while still others seek a mechanism for objective collapse. The Wigner's Friend paradox thus becomes a crucible for testing our philosophical stance on the very fabric of existence, pushing us to articulate what we mean by "being" in a quantum universe.

### 6.2 Epistemological Considerations

Beyond what is real, Wigner's friend experiment raises critical epistemological questions – questions about the nature of knowledge, how we acquire it, and its limitations. If observations are observer-dependent, and if different observers can legitimately disagree on outcomes without being "wrong," how can we reconcile these subjective observations with the pursuit of universal truths? The scientific method traditionally relies on the idea of intersubjective verifiability: experiments should yield the same results for all competent observers, leading to universally agreed-upon facts.

The extended Wigner's Friend scenarios, especially as illuminated by these no-go theorems, suggest a profound tension with this traditional scientific ideal. If "facts" are relative to an observer, then the notion of an absolute, universally accessible body of scientific knowledge becomes complicated. How do we then build a coherent scientific understanding of the universe if the very events we observe are not universally agreed upon? These questions have significant implications for both the philosophy of science and the practical conduct of scientific inquiry. They prompt us to refine our understanding of what constitutes "knowledge" in a quantum context, whether universal truths can still be sought, and how consensus might be achieved among observers who genuinely experience different quantum realities. The epistemological challenge is to bridge the gap between individual, subjective experience and the collective, objective aims of scientific understanding.

However this goal is not hopeless but will be part of a forthcoming work.

## 7. Conclusion

The "Wigner's Friend" thought experiment, conceived decades ago by Eugene Wigner, has evolved from a curious philosophical puzzle into a powerful conceptual tool for probing the very foundations of quantum mechanics. Its original formulation vividly illustrated the perplexing ambiguities surrounding the measurement problem and the enigmatic role of the observer, particularly the contentious idea that consciousness might play a part in collapsing the wave function. It exposed the deep divides between interpretations of quantum mechanics, from the initial Copenhagen view to the many-branched universe of the Many-Worlds interpretation.

Crucially, the advent of extended Wigner's Friend scenarios and their experimental approximations has reinvigorated this debate, transforming it from purely abstract speculation into an area of active theoretical and empirical investigation. These modern iterations, involving nested observers and intricate quantum correlations, have intensified the paradoxes, leading to profound insights such as the various no-go theorems, which suggests an inherent tension between the universality of quantum mechanics and the existence of a single, objective reality shared by all observers.

The implications of these extended interpretations are far-reaching. They fundamentally challenge classical intuitions about realism, objectivity, and the nature of physical events, suggesting that "facts" themselves may be relative to the observer. This perspective has bolstered perspectival interpretations such as ConSol and necessitates a re-evaluation of the status of reality in a quantum universe. Moreover, these scenarios are not just philosophical exercises; they offer critical insights for quantum information theory, informing our understanding of entanglement, decoherence, and the very principles guiding quantum computation.

As theoretical frameworks become more refined, Wigner's friend and its extensions will undoubtedly remain pivotal in exploring the ultimate nature of reality. They compel us to confront our most cherished assumptions about what constitutes a definitive event, a shared reality, and the boundaries between the subjective and the objective.

**Funding**: None
**Institutional Review Board statement**: not applicable
**Informed Consent Statement**: not applicable
**Data Availability Statement**: not applicable
**Conflicts of Interest:** The author declares no conflict of interest

.